# A sensitivity analysis of research institutions' productivity rankings to the time of citation observation[1]

*Giovanni Abramo*[a,b,*], *Tindaro Cicero*[b], *Ciriaco Andrea D'Angelo*[b]

[a] Institute for System Analysis and Computer Science (IASI-CNR)
National Research Council of Italy

[b] Laboratory for Studies of Research and Technology Transfer
School of Engineering, Department of Management
University of Rome "Tor Vergata"

**Abstract**

One of the critical issues in bibliometric research assessments is the time required to achieve maturity in citations. Citation counts can be considered a reliable proxy of the real impact of a work only if they are observed after sufficient time has passed from publication date. In the present work the authors investigate the effect of varying the time of citation observation on accuracy of productivity rankings for research institutions. Research productivity measures are calculated for all Italian universities active in the hard sciences in the 2001-2003 period, by individual field and discipline, with the time of the citation observation varying from 2004 to 2008. The objective is to support policy-makers in choosing a citation window that optimizes the tradeoff between accuracy of rankings and timeliness of the exercise.

**KEYWORDS**
*Research evaluation; bibliometrics; citation window; productivity; universities; rankings*

---

**\* Corresponding author:** Dipartimento di Ingegneria dell'Impresa, Università degli Studi di Roma 'Tor Vergata', Via del Politecnico 1, 00133 Rome - ITALY, tel/fax +39 06 72597362, giovanni.abramo@uniroma2.it

# 1. Introduction

There are an ever growing number of nations that now carry out regular evaluation exercises of their overall research systems. The inroad of bibliometric indicators to integrate peer-review in evaluation exercises, has been made possible by continuous advancement of bibliometric techniques

The advantage of bibliometrics with respect to classic peer review rests not so much in greater effectiveness at evaluating single research outputs, as in the possibility of measuring productivity by evaluating all the publications (indexed in qualified sources such as the Web of Science or Scopus), which are highly representative of the entire research output, even if only in the hard sciences. This certainly does not rend bibliometric evaluation perfect, but in these disciplines definitely makes it better than peer-review in terms of robustness, validity, functionality, costs and time of execution as showed in Abramo and D'Angelo, 2011.

Thus there is no surprise that the Australian government chose to use bibliometric methods alone for comparative evaluation of universities in the hard sciences, in the assessment framework called "Excellence in Research for Australia" (ERA, 2010). Other nations, such as Great Britain with the upcoming Research Evaluation Framework (REF 2009), and Italy with the announced "Quality Assessment of Research" (VQR, 2011), have made more modest changes, with compromise solutions involving "informed peer review", in which peer reviewers can draw on bibliometric indicators in forming their judgments of research products.

The seemingly unstoppable expansion of bibliometric evaluation stimulates scholars to continuously refine the methods of application and resolve the inherent limits of techniques. One of the concerns is citation as an indicator of impact of scientific output. There is a shared opinion that citation count can be considered a reliable proxy of real impact of a work only if observed after sufficient time has passed from the date of publication (Glänzel et al., 2003). This gives rise to a difficult balance between the needs of policy-makers and research institution managers to receive performance rankings as close as possible to the observed time-period and the need for a sufficient window for the citations to accumulate and provide a robust indicator of impact. The question of the most appropriate citation window length is then an important one. Rousseau (1988) noted that in certain fields (e.g. mathematics-related), the standard bibliometric time horizon is greater than in others: for correct evaluation of impact of a work in mathematics the citation window should be more than three years. A subsequent study by Adams (2005) concludes that citations received 1 and 2 years after publication "might be useful as a forward indicator of the long-term quality of research publications". In a previous work Abramo et al. (2011a) attempted to provide quantitative meaning to "sufficient", analyzing citation speeds and patterns for Italian publications under windows of various lengths of time. The results confirmed previous literature indicating that different fields show different citation patterns and that citation speed is quite different for clusters of disciplines. However, with the sole exception of Mathematics, the authors argue that a time lapse of two or three years between date of publication and citation observation appears a sufficient guarantee of robustness in impact indicators for single research products. A greater time lag would offer greater accuracy, but with ever decreasing incremental effect and with further delay in carrying out the evaluation.

In the present work the authors propose a step forward, investigating the effect of citation window length not on the accuracy of measuring single publication impact, but rather in determining the productivity rankings of overall research institutions. To do this we measure the research productivity of all Italian universities active in the hard sciences over the period 2001-2003, at the level of their individual research fields and disciplines, with the time of citation observation varying from year 2004 to 2008. A first important step is to conduct a sensitivity analysis of research productivity rankings to the time of citation observation. Next we provide an indication of the extent of error in the ranks as citation window shortens, by field and discipline.

The following section describes the dataset and methodologies used in the analyses. Section 3 of the paper presents the results from the elaborations and a final section summarizes the results and provides the authors' considerations on policy implications.

## 2. Dataset and methodology

Because of the different intensity of publication and citation across scientific fields, bibliometric comparison of research institution performance must be conducted at the level of individual field. Thus it is necessary to identify the research fields for the personnel in research institutions and then compare the productivity of researchers from the same fields. In the hard sciences, the research staff of Italian universities are classified in 205 fields (named scientific disciplinary sectors, SDSs[2]) grouped into nine disciplines (named university disciplinary areas, UDAs[3]). We assume the SDS as unit of analysis: measures of productivity are applied to the research staff of every university active in the SDS. Data on staff members of each university and their SDS classifications are extracted from the database on Italian university personnel, maintained by the Ministry for Universities and Research[4]. The bibliometric dataset used to measure output of research is extracted from the Italian Observatory of Public Research (ORP)[5], a database developed and maintained by the authors and derived under license from the Thomson Reuters' Web of Science (WoS). Beginning from the raw data of the WoS and applying a complex algorithm for reconciliation of the author's affiliation and disambiguation of the true identity of the authors, each publication (article, article review and conference proceeding) is attributed to the university scientist or scientists that produced it. The procedures involved are fully explained in D'Angelo et al., 2010. The authors note that there are certain limitations in WoS coverage (Van Leeuwen et al., 2001) that must be taken into account in interpreting results.

For the current study, to ensure the representativity of publications as proxy of research output, the field of observation was limited to those SDSs (184 in all) where at least 50% of

---

[2] Complete list accessible at http://attiministeriali.miur.it/UserFiles/115.htm, last accessed on November 11, 2011.
[3] Mathematics and computer sciences; physics; chemistry; earth sciences; biology; medicine; agricultural and veterinary sciences; civil engineering; industrial and information engineering
[4] http://cercauniversita.cineca.it/php5/docenti/cerca.php, last accessed on November 11, 2011
[5] www.orp.researchvalue.it, last accessed on November 11, 2011

Italian university scientists produced at least one publication in the observed period. The dataset thus composed consists of 79,715 publications authored by a total of 32,377 Italian university scientists, in 184 SDSs: Table 1 shows the distribution of publications among the 184 SDSs and 9 UDAs.

| UDA | N° of SDSs | N° of Universities | N° of research staff | N° of publications* |
|---|---|---|---|---|
| Mathematics and computer science | 9 | 58 | 2,901 | 7,112 |
| Physics | 8 | 57 | 2,484 | 15,519 |
| Chemistry | 12 | 58 | 3,057 | 16,502 |
| Earth sciences | 12 | 48 | 1,253 | 2,665 |
| Biology | 19 | 63 | 4,752 | 18,146 |
| Medicine | 47 | 54 | 10,035 | 34,532 |
| Agricultural and veterinary sciences | 28 | 40 | 2,525 | 4,983 |
| Civil engineering | 7 | 45 | 1,166 | 2,130 |
| Industrial and information engineering | 42 | 60 | 4,204 | 15,628 |
| Total | 184 | 66 | 32,377 | 79,715 |

*Table 1: Numbers of SDSs, universities, research staff and publications for the Italian academic system, by UDA; data 2001-2003*
\* A publication is assigned to multiple UDAs if co-authored by researchers falling in different UDAs.

Rather than considering simple output to calculate the productivity of a university researcher we consider the actual outcome, or "impact", of the research in the researcher's scientific field. As proxy of outcome we adopt the number of citations for the researcher's publications. All the noted limits concerning citations as proxy of impact apply (Pendlebury, 2009). Researchers belonging to a particular scientific field may also publish outside that field: a statistician may publish in a medical science journal or a physicist in bibliometrics (a famous example being the physicist Jorge E. Hirsch, developer of the h-index). For this reason we standardize the citations for each publication accumulated at December 31 of each year for citations windows 2004-2008, with respect to the median[6] for the distribution of citations for all the Italian publications of the same year and the same WoS subject categories. This standardized impact for a publication is called the Article Impact Index (AII).

For a general publication in subject category $j$[7], AII observed at year $i$ is given by:

$$AII_{ij} = \frac{c_i}{Me_{ij}}$$

Where:
$c_i$ = citations received by a publication as of year $i$;
$Me_{ij}$ = median of the distribution[8] of citations received as of year $i$, for all Italian publications of the same year and subject category $j$.

---

[6] As frequently observed in literature (Lundberg, 2007), standardization of citations with respect to median value rather than to the average is justified by the fact that distribution of citations is highly skewed in almost all disciplines.
[7] For publications in multidisciplinary journals the AII is calculated as a weighted average of the standardized values for each subject category.
[8] Publications without citations are excluded from calculation of the median.

Thus we proceed to measurement of the impact indicator Scientific Strength[9] (SS), for each university and SDS. This is given by the sum of the publications produced by the researchers in a university SDS[10], each weighted for AII. For a generic SDS of a generic university:

$$SS = \sum_{k=1}^{n} AII_k$$

Where:
*n* is the number of publications of researchers of the SDS of the university in the period of observation.

At this point we can calculate productivity (p) of an SDS as the ratio of Scientific Strength to the number of research staff (RS) in the SDS:

$$p = \frac{SS}{RS}$$

Since national research assessment exercises generally elaborate university rankings at the level of discipline (i.e. UDA), we calculate productivity (*P*) of a general UDA of a general university:

$$P = \sum_{w=1}^{n} \frac{p_w}{\overline{p_w}} \frac{RS_w}{RS} = \sum_{w=1}^{n} \frac{SS_w}{\overline{p_w} \, RS}$$

where:
  $p_w$ = productivity of the SDS *w*
  $\overline{p_w}$ = average productivity of national universities in SDS *w*
  $RS_w$ = number of scientists in SDS *w*
  $RS$ = number of scientists in the UDA
  $n$ = number of SDSs in the UDA

Through this procedure, first calculating productivity at the SDS level, then standardizing to national average and weighting for the relative size of the SDS in the UDA, we take account of the varying intensity of publication and citation for the SDSs, avoiding the typical distortion of measures at the aggregate level of discipline (Abramo et al., 2008). There remain the limits concerning possible differences in availability of production factors other than labor across universities, though in the Italian case the assumption of uniform distribution is acceptable (Abramo et al., 2011b). In particular we assume a uniform distribution of capital per research staff, since in Italy the large part of financial resources is equally allocated by government to satisfy the needs of each university in function of its size. The potential greater availability of funds per staff unit in a university is thus due to its capacity to acquire such funds on a competitive basis. Greater output deriving from greater availability of funds is thus the result of merit and not of any other comparative advantages.

---

[9] SS is similar to the "crown indicator" of CWTS and the "total field normalized citation score" of the Karolinska Institute. The differences are: i) we standardize citations of single publications and not of scientific portfolio of researchers/institutions; ii) we standardize by the Italian median rather than the world average.
[10] Publications are assigned to universities and SDSs. In case of co-authorship of scientists belonging to different universities and different SDSs there are multiple counting of the same publication.

In any case such limits would not unbalance our analysis, since the objective is to determine variations in ranking and not the absolute value of productivity.

## 3. Results and analysis

### 3.1 Variations in rankings

The first objective is to analyze how university rankings vary with variation of the citation window. Thus in this section we show the variation of productivity rankings (period 2001-2003) with variation in the year for counting citations, from 2004 to 2008.

We begin by showing the procedure for calculating the productivity of a university in a specific discipline, taking the example of the University of Naples 'Federico II' in the UDA of Mathematics and computer science (Table 2), with citations counted at the close of 2008. In this UDA there are 162 research staff divided amongst 8 SDSs. In the triennium 2002-2003 they produced 281 publications. The absolute values for productivity (p) of each SDS are shown in Column 5. Column 6 shows the average value for productivity of all Italian universities ($\bar{p}$). The last column shows the standardized and weighted values: summation of the values for all SDSs of the UDA indicates the productivity for the UDA as equal to 0.576.

| SDS | RS | Publications | SS | p (SS/RS) | $\bar{p}$ | $\frac{p}{\bar{p}} \cdot \frac{RS}{162}$ |
|---|---|---|---|---|---|---|
| MAT/02 – Algebra | 13 | 19 | 4.128 | 0.318 | 0.432 | 0.059 |
| MAT/03 - Geometry | 31 | 50 | 7.810 | 0.252 | 0.633 | 0.076 |
| MAT/05 – Mathematics (analysis) | 60 | 93 | 33.791 | 0.563 | 0.922 | 0.226 |
| MAT/06 - Probability and statistics | 5 | 14 | 3.840 | 0.768 | 0.624 | 0.038 |
| MAT/07 - Mathematical physics | 23 | 50 | 13.973 | 0.608 | 0.884 | 0.098 |
| MAT/08 - Numerical analysis | 13 | 17 | 6.107 | 0.470 | 1.062 | 0.035 |
| MAT/09 - Operations research | 3 | 8 | 1.011 | 0.337 | 0.875 | 0.007 |
| INF/01 - Informatics | 14 | 30 | 4.996 | 0.357 | 0.841 | 0.037 |
| Total | 162 | 281 | | | | 0.576 |

*Table 2: Productivity of University of Naples "Federico II" for the SDSs in Mathematics and computer science; data 2001-2003, observed 31/12/2008*

Comparison of all values calculated with the same method for all the Italian universities active in the UDA provides the rankings list for our analysis. With variation in the moment of observing citations there will obviously be variation in the productivity ratings, and thus in the rankings. Figure 1 again presents the case of UDA Mathematics and computer science: the x-axis shows identification numbers for the universities and the y-axis shows the range of variation for the five bibliometric rankings obtained for the years 2004 to 2008. There are a total of 58 universities active in this discipline. Of these, 95% experience at least one shift in rankings over the five scenarios considered. The university with greatest variability is the Free University of Bolzano (ID 12), which changes from 43$^{rd}$ ranked in the 2005 evaluation to 12$^{th}$ in the 2008 list. The University of Venice "Ca' Foscari" (ID 24) shows the inverse trend, shifting from 5$^{th}$ place in 2004 to 24$^{th}$ place in the 2008 list. Other institutions showing notable shifts are the University of Sannio (ID 26), which drops 14 positions between the 2004 and 2008 rankings; the University of Basilicata (ID 33), which betters 12 positions and the University of Camerino (ID 40) which improves 15.

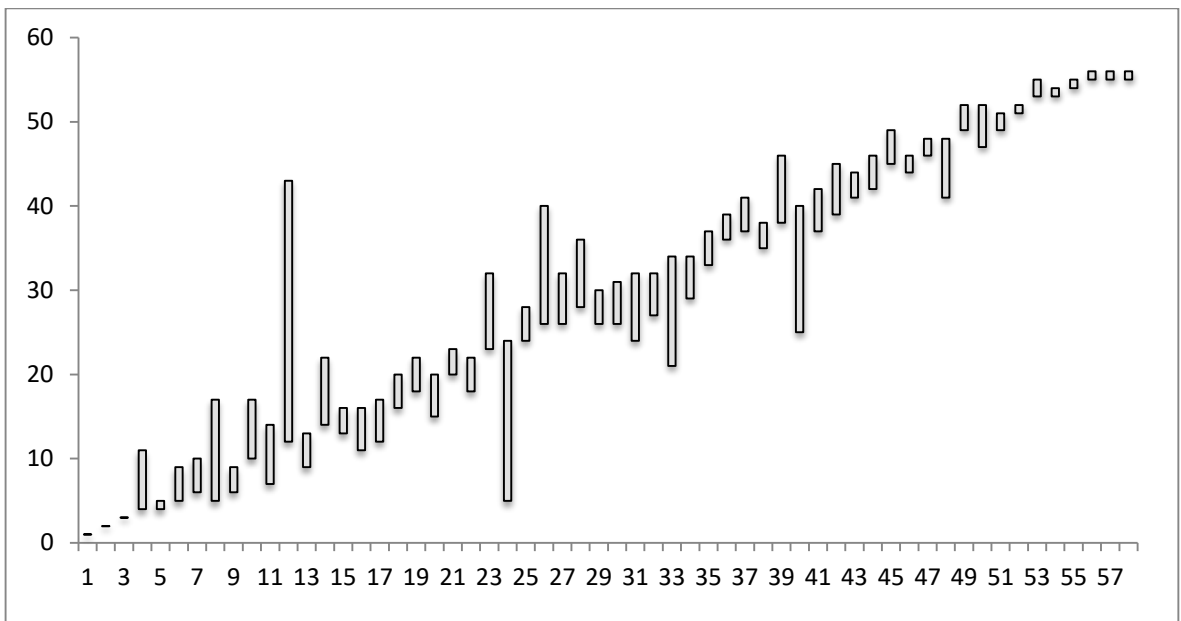

*Figure 1: Range of variation of 2001-2003 productivity rankings of Italian universities in UDA Mathematics and computer science, based on citations observed at the end of each year from 2004 to 2008.*

In the next section we provide a deeper analysis of the variations between rankings from 2008 citations and rankings from citations of preceding years.

### 3.2 Distribution of differences in rankings

Assuming the 2008 rankings as a benchmark, Figure 2 presents the shifts in rank with variation in the year of citation count, for the example of Chemistry.

Comparison of the 2004 and 2008 rankings lists shows an asymmetric distribution of rank shifts with a fairly long right tail and a peak for a one rank shift. In following years, distributions of frequencies for variations concentrate to the left with the right tail progressively shorter. Dispersion of differences drops notably from 2004 to 2005: standard deviation for rank shifts drops from 2.955 to 1.974 (Table 3). By 2007, more than half the universities show no shift in rank compared to the 2008 benchmark (median = 0).

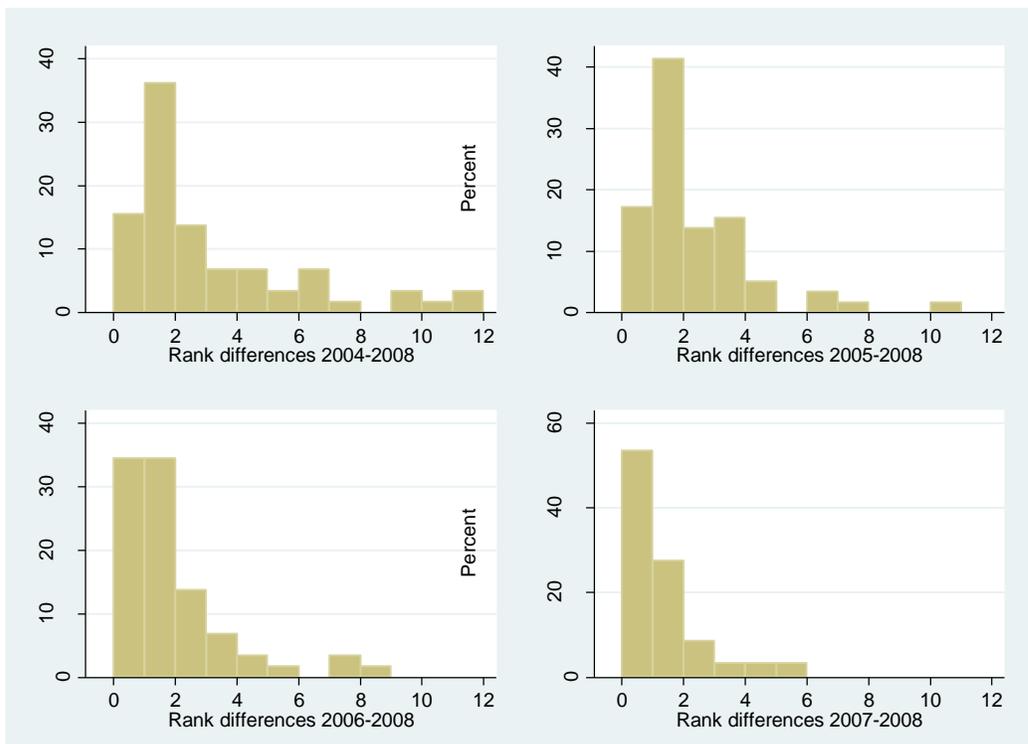

*Figure 2: Distributions of differences between 2008 productivity rankings in Chemistry and rankings from citations in previous years.*

| Descriptive statistics | 2004 vs 2008 | 2005 vs 2008 | 2006 vs 2008 | 2007 vs 2008 |
|---|---|---|---|---|
| Mean | 2.707 | 1.879 | 1.448 | 0.896 |
| Median | 1 | 1 | 1 | 0 |
| Standard dev. | 2.955 | 1.974 | 1.884 | 1.398 |
| Skewness | +1.527 | +2.280 | +2.129 | +2.129 |
| Kurtosis | +4.605 | +9.869 | +7.804 | +7.522 |

*Table 3: Descriptive statistics for differences between 2001-2003 university productivity rankings in Chemistry (benchmark citations observed 2008) compared to rankings from citations in previous years*

The same analysis was repeated for all the UDAs: Table 4 shows the descriptive statistics for the variations in ranking of universities in each UDA. Under variation in year of observation, the percentage of universities showing a shift in rank reaches a high of 98%: this case is seen in the two UDAs of Industrial and information engineering and Earth science. These two UDAs, together with Mathematics, also show the highest average shift (column 4). The most extreme case of variability is seen in Industrial and information engineering, where one of the universities shifts 39 positions in ranking (last column) between the worst and best years (2004, 2008). In Mathematics there is one university (out of 58 total) that shifts 31 positions, while in Earth Sciences the most extreme case is a university (of 48 total) that shifts 26 positions.

| UDA | Total Universities | % change | average | median | Std Dev. | Max ranking variation |
|---|---|---|---|---|---|---|
| Mathematics and computer science | 58 | 95% | 2.750 | 1.875 | 2.691 | 31 |
| Physics | 57 | 89% | 1.184 | 0.750 | 1.210 | 11 |
| Chemistry | 58 | 95% | 1.732 | 1.250 | 1.534 | 12 |
| Earth sciences | 48 | 98% | 2.141 | 1.500 | 2.055 | 26 |
| Biology | 63 | 83% | 1.452 | 0.750 | 1.623 | 16 |
| Medicine | 54 | 89% | 0.944 | 0.750 | 1.012 | 8 |
| Agricultural and veterinary sciences | 40 | 83% | 1.269 | 0.750 | 1.475 | 13 |
| Civil engineering | 45 | 96% | 1.683 | 1.000 | 1.571 | 11 |
| Industrial and information engineering | 60 | 98% | 3.258 | 2.000 | 3.558 | 39 |

*Table 4: Descriptive statistics by UDA for variation in university productivity rankings from 2008 citations with rankings from citations of preceding years*

A similar analysis was conducted at the level of SDS, to detect potential differences across SDSs and with respect to overall UDAs. As an example we show descriptive statistics for the SDSs in the UDA of Mathematics and computer science (Table 5). As can be seen, the variations by SDS are slightly greater than those for the overall UDAs.

| UDA | Tot. Universities | % change | average | median | Std Dev. | Max ranking variation |
|---|---|---|---|---|---|---|
| MAT/01 – Mathematical logic | 20 | 55% | 0.500 | 0.250 | 0.644 | 4 |
| MAT/02 - Algebra | 39 | 97% | 2.128 | 1.750 | 1.658 | 27 |
| MAT/03 - Geometry | 50 | 96% | 2.130 | 1.250 | 1.792 | 14 |
| MAT/05 – Mathematics (analysis) | 53 | 91% | 2.042 | 1.750 | 1.730 | 13 |
| MAT/06 - Probability and statistics | 37 | 100% | 1.709 | 1.500 | 1.205 | 9 |
| MAT/07 - Mathematical physics | 45 | 98% | 1.678 | 1.250 | 1.225 | 9 |
| MAT/08 - Numerical analysis | 44 | 98% | 2.534 | 2.625 | 1.572 | 13 |
| MAT/09 - Operations research | 34 | 97% | 2.228 | 1.750 | 1.646 | 18 |
| INF/01 - Informatics | 45 | 96% | 3.006 | 2.000 | 2.895 | 30 |

*Table 5: Descriptive statistics by SDS (example of Mathematics and computer science) for variation in university productivity rankings from 2008 citations with rankings from citations of preceding years*

Returning to the UDA level, the overall observation is that, with the exceptions of very few universities, the rankings remain substantially stable over the five scenarios prepared. This is confirmed by Table 6 showing the Spearman correlation coefficients between the rankings from evaluation at each year end and the last set of rankings, at end of 2008, which provide the analysis benchmark.

| UDA | Rank_2004 | Rank_2005 | Rank_2006 | Rank_2007 |
|---|---|---|---|---|
| Mathematics and computer science | 0.934 | 0.949 | 0.987 | 0.989 |
| Physics | 0.984 | 0.992 | 0.997 | 0.998 |
| Chemistry | 0.972 | 0.987 | 0.990 | 0.995 |
| Earth sciences | 0.917 | 0.973 | 0.986 | 0.988 |
| Biology | 0.976 | 0.988 | 0.996 | 0.997 |
| Medicine | 0.986 | 0.993 | 0.996 | 0.999 |
| Agricultural and veterinary sciences | 0.961 | 0.986 | 0.979 | 0.993 |
| Civil engineering | 0.954 | 0.968 | 0.987 | 0.995 |
| Industrial and information engineering | 0.863 | 0.956 | 0.974 | 0.989 |
| *Average* | *0.950* | *0.977* | *0.988* | *0.994* |

*Table 6: Correlation of 2001-2003 productivity rankings from 2008 citations with rankings from citations*

*of preceding years*

The correlation between the 2004 and 2008 productivity rankings[11] is always greater than 0.9, except for Industrial and information engineering ($\rho=0.863$). When observations are taken in subsequent years their correlation to 2008 rankings rises but in constantly decreasing manner: with the evaluation conducted in 2005 the correlation to benchmark for Industrial and information has already corrected to 0.956. The data thus suggest that the rankings lists tend to stabilize very rapidly, as soon as within one year from the terminal date of the period under evaluation.

Table 7 shows comparisons between 2004 and 2008 only for rankings in all UDAs, indicating the percentages of universities with no shift and with shifts less than or equal to three positions. In Industrial and Information Engineering only 3% of universities escape without rank shift. This area, and also Mathematics and computer science and Earth science, show distinctly greater variability compared to other UDAs. On the other hand, the Agricultural and veterinary science UDA has the greatest percentage of universities showing no change (28%) and Medicine has the most with shift less than or equal to 3 positions (87%).

| UDA | N° of Universities | No change | Change ≤ 3 rank shifts |
|---|---|---|---|
| Mathematics and computer sciences | 58 | 9% | 64% |
| Physics | 57 | 18% | 81% |
| Chemistry | 58 | 16% | 72% |
| Earth sciences | 48 | 8% | 71% |
| Biology | 63 | 27% | 73% |
| Medicine | 54 | 22% | 87% |
| Agricultural and veterinary sciences | 40 | 28% | 85% |
| Civil engineering | 45 | 22% | 76% |
| Industrial and information engineering | 60 | 3% | 57% |

*Table 7: Comparison of differences in university productivity rankings, by UDA, based on citations observed 2004 and 2008 for publication period 2001-2003.*

The results seen here lend confirmation to a previous study by Abramo et al. (2011a). The 2011 study, examining publications in Mathematics and Engineering, showed a generally constant trend for increase in citations, with a peak in the final year of observation (2008). In other disciplines the citation patterns generally peaked within two or three years after publication, suggesting that for these UDAs there would be even less variation in any university rankings from bibliometric evaluation exercises.

In the next section we examine a realistic aspect of the question of rank variations by subdividing the universities into performance classes and analyzing the average shifts in rank for these classes.

**3.3 Quartiles variation of universities productivity**

In most real-world assessment exercises the performance profile of universities is expressed in quartiles, so we classify Italian universities into four classes by productivity, assigning values of 4, 3, 2 and 1, corresponding to first, second, third and fourth quartiles for the productivity distribution in the UDA. As previous, the analysis takes 2008 as benchmark and presents four other scenarios for date of observation. Table 8 shows the average values of class shift by UDA. The differences between 2004 and 2008 repeat the data patterns seen previously: Industrial and information engineering shows the highest value of average variation in class (0.400); Physics shows the lowest value (0.105) and by 2007, Physics and Medicine register no further shifts in class. By 2007, the greatest shifts remaining are in Mathematics and computer science (0.138) and Civil engineering (0.133); Civil engineering shows this same 0.133 average shift for final three scenarios in succession.

---
[11] This means the productivity rankings from observations at 31/12 of these years.

| UDA | 2004 vs 2008 | 2005 vs 2008 | 2006 vs 2008 | 2007 vs 2008 |
|---|---|---|---|---|
| Mathematics and computer science | 0.345 | 0.241 | 0.207 | 0.138 |
| Physics | 0.105 | 0.035 | 0.035 | 0.000 |
| Chemistry | 0.241 | 0.138 | 0.103 | 0.103 |
| Earth sciences | 0.333 | 0.167 | 0.042 | 0.042 |
| Biology | 0.159 | 0.095 | 0.063 | 0.063 |
| Medicine | 0.148 | 0.074 | 0.037 | 0.000 |
| Agricultural and veterinary sciences | 0.150 | 0.100 | 0.050 | 0.050 |
| Civil engineering | 0.178 | 0.133 | 0.133 | 0.133 |
| Industrial and information engineering | 0.400 | 0.267 | 0.133 | 0.100 |

*Table 8: Average quartiles differences in university productivity rates, by UDA, comparing rates from 2008 citations and previous years.*

It is informative to also examine the numbers of outliers, or universities with shifts of two or three productivity quartiles[12]: Table 9 shows there are very few such anomalous variations. Comparing productivity rankings between 2004 and benchmark 2008, only four universities show variation of two or three quartiles: two of these cases are in Earth sciences; one is in Mathematics and one in Industrial and information engineering. By 2005 there are only two universities that shift two or three classes from benchmark: one each in Earth sciences and Mathematics. From 2006 on there are no productivity shifts of more than a quartile.

| UDA | 2004 vs 2008 | 2005 vs 2008 | 2006 vs 2008 | 2007 vs 2008 |
|---|---|---|---|---|
| Mathematics and computer science | 1 | 1 | 0 | 0 |
| Physics | 0 | 0 | 0 | 0 |
| Chemistry | 0 | 0 | 0 | 0 |
| Earth sciences | 2 | 1 | 0 | 0 |
| Biology | 0 | 0 | 0 | 0 |
| Medicine | 0 | 0 | 0 | 0 |
| Agricultural and veterinary sciences | 0 | 0 | 0 | 0 |
| Civil engineering | 0 | 0 | 0 | 0 |
| Industrial and information engineering | 1 | 0 | 0 | 0 |

*Table 9: Number of universities showing two or three quartiles variations in productivity rates, by UDA*

---

[12] The maximum shift is three quartiles: this can occur if a university places in first category for the 2008 benchmark but last category in some previous year, or vice versa.

A final question for investigation is whether specific classes of universities show greater (or lesser) variability in rank, in particular whether "top" (or "bottom") universities tend to experience such shifts. We define the "top universities" for each UDA as those that place above 80th percentile in the 2008 benchmark productivity ratings. Our particular interest is if top universities show less variability than others. We apply the NPC Test[13], with null hypothesis $h_0$: $average(top) = average(no\_top)$, where "average" is the average maximum difference of rank between the benchmark year of 2008 and previous years. The results show that there are only two UDAs where "2008 top" universities average a lower number of rank shifts than other universities (Table 10). In all the other UDAs there are no significant differences in variations between top universities and the others.

| UDA | p-value (top universities vs all others) |
|---|---|
| Mathematics and computer science | 0.847 |
| Physics | 0.295 |
| Chemistry | 0.071*(<) |
| Earth sciences | 0.148 |
| Biology | 0.152 |
| Medicine | 0.202 |
| Agricultural and veterinary sciences | 0.023**(<) |
| Civil engineering | 0.203 |
| Industrial and information engineering | 0.332 |
| Combined test F | 0.019**(<) |

*Table 10: NPC test comparing average values of maximum rank shift for top universities and all other universities*

## 4. Conclusions

The current diffusion of evaluation exercises for national research systems is linked to development of bibliometric techniques, either in integration or complete substitution of the classic peer review methods. The rapidity and frequency for the conduct of evaluations thus also depends on developments in bibliometric techniques. Scholars strive to deal with the intrinsic limits of the technique. One of the issues, concerning the citation indicator is that citation counts can be considered a reliable proxy of real impact of a work only if observed at sufficient distance in time from the date of publication. This gives rise to conflict between the need for evaluations to be conducted as quickly as possible after the period of interest and the need for accuracy and robustness in the rankings of institutional performance.

The current work is intended to provide useful information concerning this trade-off, which will serve policy-makers in their choices for timing of evaluation exercises. Taking the case of Italian universities, it provides analysis of the sensitivity of productivity rankings to length of citation window. For the evaluation period 2001-2003, the results show substantial stability in performance rankings as citation window varies from 2004 to 2008, with the exception of a very limited number of universities showing exceptional changes. Given the 2008 evaluation as benchmark, the correlation to rankings taken as early as 2004 is already greater than 0.9, with a sole exception for the discipline of Industrial and information engineering (0.863). This discipline shows greater variability than all the others, with next greatest variability in Earth sciences and Mathematics. In another analysis we subdivide the universities by quartile according to their productivity: comparing the quartile rankings for the extreme years (2004 and 2008), only four universities show shifts of more than one quartile – two of these cases are in Earth sciences, one in Mathematics and one in Industrial and information engineering. Over the full term of the five windows considered there is general stability in rankings, with Agricultural and veterinary science and Medicine being the disciplines that are most stable of all.

---

[13] Non parametric combination of dependent permutation tests (Pesarin, 2001)

In summary, the study shows that the tradeoff between accuracy and unwanted delay in carrying out evaluation is much less dramatic than might have been expected. The accuracy of bibliometric assessment for university research productivity seems quite acceptable within one year from the close of a given three-year period. As a further logical conclusion, the trade-off would be even less for evaluation of productivity over periods longer than three years.

From a previous study we have seen that accurate measurement of impact of a single publication requires a citation window length of two to three years (Abramo et al. 2011a), When it comes to comparing the performance ranking of research institutions, it seems sufficient to count citations one year after the upper limit of a three-year production period.. The results here stimulate the question of what citation window is necessary for evaluation at other levels, particularly for individual scientists. A research project on the subject has been concluded during the review of this manuscript. Hopefully, the results will be made public soon.

## References

Abramo G., D'Angelo C.A., Di Costa F., (2008). Assessment of sectoral aggregation distortion in research productivity measurements. *Research Evaluation,* 17(2), 111-121.

Abramo G., Cicero T., D'Angelo C.A. (2011a). Assessing the varying level of impact measurement accuracy as a function of the citation window length, *Journal of Informetrics*, DOI:10.1016/j.joi.2011.06.004

Abramo G., Cicero T., D'Angelo C.A., (2011b). A field-standardized application of DEA to national-scale research assessment, *Journal of Informetrics*, DOI: 10.1016/j.joi.2011.06.001

Abramo G., D'Angelo C.A., (2011). Evaluating research: from informed peer review to bibliometrics, *Scientometrics*, 87(3), 499-514.

Adams J., (2005). Early citation counts correlate with accumulated impact. *Scientometrics,* 63(3), 567-581.

D'Angelo C.A., Giuffrida C., Abramo G., (2011). A heuristic approach to author name disambiguation in large-scale bibliometric databases, *Journal of the American Society for Information Science and Technology*, 62(2), 257–269.

ERA, (2010). The Excellence in Research for Australia (ERA) Initiative,

http://www.arc.gov.au/era/ last accessed on November 11, 2011.

Glänzel W., Schlemmer B., Thijs B., (2003). Better late than never? On the chance to become highly cited only beyond the standard bibliometric time horizon. S*cientometrics,* 58(3), 571-586.

Lundberg J., (2007). Lifting the crown-citation z-score. *Journal of Informetrics*, 1(2), 145–154.

Pendlebury D.A., (2009). The use and misuse of journal metrics and other citation indicators. *Archivum Immunologiae et Therapiae Experimentalis*, 57(1), pp. 1-11.

Pesarin F., (2001). *Multivariate Permutation tests: with application in Biostatistics*. Ed. John Wiley & Sons, Chichester-New York.

RAE, (2008). Research Assessment Exercise – 2008. Panel criteria and working methods. http://www.rae.ac.uk/pubs/2006/01/ last accessed on November 11, 2011.

REF, (2009). Report on the pilot exercise to develop bibliometric indicators for the Research Excellence Framework. http://www.hefce.ac.uk/pubs/hefce/2009/09_39/ last accessed on November 11, 2011.

Rousseau R., (1988), Citation distribution of pure mathematics journals. In: *Egghe, L., Rousseau, R. (Ed.) Informetrics,* Belgium: Diepenbeek, *87/88*, 249-262, *Proceedings 1st International Conference on Bibliometrics and Theoretical Aspects of Information Retrieval*.

VQR, (2011). *Linee guida VQR 2004-2008, http://www.civr.miur.it/vqr_decreto.html* (last accessed on November 11, 2011).

Van Leeuwen T.N., Moed H.F., Tijssen R.J.W., Visser M.S., van Raan A.F.J., (2001). Language biases in the coverage of the Science Citation Index and its consequences for international comparisons of national research performance, *Scientometrics*, 51(1) 35-346.